ORIGINAL ARTICLE

# Differentiating Radiation Necrosis and Metastatic Progression in Brain Tumors Using Radiomics and Machine Learning

*Elahheh Salari, PhD,\* Haitham Elsamaloty, MD,† Aniruddha Ray, PhD,‡§ Mersiha Hadziahmetovic, MD,\* and E. Ishmael Parsai, PhD\**

**Objectives:** Distinguishing between radiation necrosis (RN) and metastatic progression is extremely challenging due to their similarity in conventional imaging. This is crucial from a therapeutic point of view as this determines the outcome of the treatment. This study aims to establish an automated technique to differentiate RN from brain metastasis progression using radiomics with machine learning.

**Methods:** Eighty-six patients with brain metastasis after they underwent stereotactic radiosurgery as primary treatment were selected. Discrete wavelets transform, Laplacian-of-Gaussian, Gradient, and Square were applied to magnetic resonance post-contrast T1-weighted images to extract radiomics features. After feature selection, dataset was randomly split into train/test (80%/20%) datasets. Random forest classification, logistic regression, and support vector classification were trained and subsequently validated using test set. The classification performance was measured by area under the curve (AUC) value of receiver operating characteristic curve, accuracy, sensitivity, and specificity.

**Results:** The best performance was achieved using random forest classification with a Gradient filter (AUC = $0.910 \pm 0.047$, accuracy $0.8 \pm 0.071$, sensitivity = $0.796 \pm 0.055$, specificity = $0.922 \pm 0.059$). For, support vector classification the best result obtains using wavelet_HHH with a high AUC of $0.890 \pm 0.89$, accuracy of $0.777 \pm 0.062$, sensitivity = $0.701 \pm 0.084$, and specificity = $0.85 \pm 0.112$. Logistic regression using wavelet_HHH provides a poor result with AUC = $0.882 \pm 0.051$, accuracy of $0.753 \pm 0.08$, sensitivity = $0.717 \pm 0.208$, and specificity = $0.816 \pm 0.123$.

**Conclusion:** This type of machine-learning approach can help accurately distinguish RN from recurrence in magnetic resonance imaging, without the need for biopsy. This has the potential to improve the therapeutic outcome.

**Key Words:** brain metastasis progression, machine learning, radiation necrosis, radiomics

Brain metastasis is generally difficult to manage, and they normally decrease a patient's quality of life considerably exhibiting signs and symptoms like headaches, personality changes, memory loss, seizures, etc. Different treatment techniques such as whole-brain radiation therapy, stereotactic radiosurgery (SRS), gamma knife, etc. have been used to treat brain metastasis. Presently, SRS is widely used to treat brain metastasis due to advances in the design of multi-leaf collimators, and the invention of flattening filter-free beams (FFF).[1] Unfortunately, Radiation necrosis (RN) is a common adverse effect of this treatment technique due to the delivery of a high dose of radiation to the tumor in one fraction or a few fractions; This technique is referred to as hypofractionated stereotactic radiotherapy. In 2015, Kohutek et al[2] reported that RN was observed in 25.8% of treated brain lesions with the SRS technique. Minniti et al[3] conducted a study on the risk of brain radionecrosis after SRS treatment and showed RN occurred in 24% of treated lesions.

Patients diagnosed with neurological symptoms usually undergo diagnostic tests, including magnetic resonance imaging (MRI).[4] Follow-up imaging is usually performed to monitor the treatment effects of radiation therapy to assess treatment responses such as complete or partial response, progressive disease, stable disease, etc. Unfortunately, RN and tumor progression appear similar on conventional MRI sequences due to disruption of the blood-brain barrier.[5] Therefore, a crucial task for clinicians and radiologists is the ability to differentiate between RN and tumor recurrence from these MRI images.

Recently, radiomics has been used in medicine including radiotherapy to predict or evaluate treatment outcomes. Different studies have been conducted using radiomics to analyze and evaluate follow-up images.[6–8] In this area, MRI plays an essential role because MR images are capable of producing superior anatomic information about the brain and other cranial structures that are clearer and more detailed than other imaging methods. Also, MRI is a noninvasive and nondestructive method to examine the tumor repeatedly to evaluate response to treatment and can, therefore, be integrated into therapeutic strategies.[6] Radiomics is a method of extracting mineable data from medical images and is used extensively in oncology. Extracting these data from MR imaging and linking them with underlying tissue dynamics has great potential to expand the scope of cancer imaging research. Furthermore, radiomics is a noninvasive approach that offers unlimited information that can be used for cancer detection, confirmation of prognosis, prediction of response to the treatment, and monitoring of disease status.[9] Radiomic features are examined in 2 categories: semantic (size, shape, location, etc.) and agnostic (first-order features, texture, wavelet, etc.) in another word agnostic refers to features that do not rely on prior knowledge of the underlying tissue dynamics.[10] Quantitative image features based on intensity, shape, size or volume, and texture provide information on tumor phenotype and microenvironment that is different from clinical reports, laboratory test results, and genomic or proteomic assays.[9]

This study aims to propose an automated method to differentiate RN and brain metastasis progression using radiomics. To achieve this goal, machine learning and radiomics features were used together. According to our literature review, this is the first study to evaluate the combination of different types of image processing filters such as Gradient, Square, Laplacian-of-Gaussian (LoG), and mother wavelet (Daubechies) in radiomics analysis, and 3 classifications—random forest classification (RFC), Logistic Regression, and support vector classification (SVC) on the ability of binary classification models for discriminating RN from tumor recurrence.

## METHODS

The workflow of this study includes 5 steps as follows: Clinical data collection, pre-processing, feature extraction, feature selection, and classification.

### Clinical Data Collection

After carefully reviewing all available data such as pathology test results, positron emission tomography (PET) scans, MRI, etc., only 86 out of 658 patients who had SRS or intensity-modulated Radio Surgery treatment, at least 4 sets of follow-up images, no adjuvant chemotherapy, and the large enough size of the region of interest (ROI), were selected for this study. Out of the 86 patients selected, 17 received biopsy lab results, 28 underwent PET and MRS, whereas 41 had only MRS to determine if they had RN or tumor progression. According to these findings, 38 individuals were diagnosed with RN, whereas 48 experienced a recurrence. The volume of the region of interest (ROI) needed to be large enough to capture texture information. It was shown that texture analysis can be significantly affected at ROI areas < 80 × 80 pixels but they remain unaffected at ROI areas > 180 × 180 pixels.[11] MR image sets were collected from 2 cancer centers with manufacturers (General Electric Medical System and Siemens), scanner models (signa HDxt and Aera), field strengths (3 Tesla and 1.5 Tesla), and various acquisition protocols. For instance, T1-weighted axial plane scans were performed on a Siemens Aera 1.5T MRI scanner after the administration of Prohance® gadolinium contrast agent. Acquisition parameters include a spin echo sequence with repetition time~600 msec, echo time = 8.9 msec, 4 mm slice thickness with no gap, 230 × 187 mm FOV reconstructed with a 320 × 260 matrix (0.719 mm pixel size). repetition time varies slightly between different scans, likely due to a slightly different number of images.

All MR images are clinically acceptable. All image QA tests (daily, monthly, and annually) are performed by Certified Medical Physicists.

### Pre-processing

The segmentation was done using open-source 3-dimensional (3D) slicer software (Ver 5.0.3)[12] for each MRI set of each patient slice-by-slice in the axial plane of post-contrast T1-weighted. Different follow-up images were compared with the initial MR image before treatment to evaluate tumor response to radiation and then, the target was delineated by an expert Radiologist. After this step, pre-processing filters (e.g., edge detection filters) which are commonly used for enhancing the predictive performance in radiomics studies were applied to MR images.[13–15] In the realm of image processing and computer vision applications, edge detection holds great significance. It is used to detect objects, locate boundaries, and extract features. Edge detection is a process that eliminates irrelevant data, noise, and frequencies while retaining the crucial structural characteristics of an image.[16] To detach the ROI from the background noise, we utilized edge detection filters such as discrete wavelet transformation (DWT), Gradient, and LoG filters. The square filter was also used to compare the effect of the edge detection versus non-edge detection filters on the classification.

### DWT

DWT with a total of 8 filters with 4 high pass and 4 low pass were applied to denoise images. It analyzes multifrequency phenomena localized in space, thus it can effectively extract information derived from the images.[15] A low-pass filter reduces high-frequency components while maintaining low-frequency components, resulting in blurred edges and less speckle noise in the spatial domain when applied to an image. A High-pass filter preserves high frequencies which include edges that are used for sharpening the image. Conceptually, DTW splits a signal into different frequency sub-bands. Therefore, since different types of signals contain different frequency characteristics, this behavior difference is captured in one of the frequency sub-bands. In addition, compared with Fourier transforms the wavelet transforms are very powerful because of their ability to describe any type of signal both in the time and frequency domain simultaneously. It means wavelet transform does not only tell us which frequencies are present in a signal but also at which time these frequencies have occurred. Although Fourier transform has a high resolution in the frequency domain and presents zero resolution in the time domain.[17] Usually, high-pass filters are used to obtain details from fine texture, whereas the coarse texture is obtained from low-pass filters.[10]

### Gradient Filter

One of the most common uses of gradient filter is edge detection, it detects the edges by looking for the maximum and minimum in the first derivatives of the image (Equation. 1).[16]

$$\nabla f = \begin{bmatrix} \frac{\partial f}{\partial x} \\ \frac{\partial f}{\partial y} \end{bmatrix} \qquad \text{Eq. 1}$$

Where $\frac{\partial f}{\partial x}$, and $\frac{\partial f}{\partial y}$ are the first derivatives with respect to the x and y directions. The magnitude of the gradient (Equation 2) forms the basis of various approaches for sharpening.[16]

$$|\nabla f(x, y)| = \sqrt{(\partial_x f(x, y))^2 + (\partial_y f(x, y))^2} \qquad \text{Eq. 2}$$

The gradient direction can be calculated by Equation 3.

$$\theta = \tan^{-1}(\partial_x f(x, y) / \partial_y f(x, y)) \qquad \text{Eq. 3}$$

**TABLE 1.** Selected Radiomics Features for Each Filter Include First and Second-order Features

| Filter (No. features) | Radiomics features (first and second orders) |
|---|---|
| Original (10) | 1-firstorder_TotalEnergy, 2-glcm_Imc1, 3-glcm_Imc2, 4-glcm_MCC, 5-glrlm_GrayLevelNonUniformity, 6-glrlm_RunLengthNonUniformity, 7-glszm_GrayLevelNonUniformity, 8-gldm_DependenceNonUniformity, 9-gldm_GrayLevelNonUniformity, 10-ngtdm_Coarseness |
| Gradient (8) | 1-glcm_Imc1, 2-glcm_Imc2, 3-glrlm_GrayLevelNonUniformity, 4-glrlm_RunLengthNonUniformity, 5-gldm_GrayLevelNonUniformity, 6-ngtdm_Busyness, 7-ngtdm_Coarseness, 8-ngtdm_Strength |
| Wavelet_HHH (18) | 1-firstorder_Kurtosis, 2-firstorder_TotalEnergy, 3-glcm_Idmn, 4-glcm_Idn, 5-glrlm_GrayLevelNonUniformity, 6-glrlm_LongRunHighGrayLevelEmphasis, 7-glrlm_RunLengthNonUniformity, 8-glrlm_ShortRunLowGrayLevelEmphasis, 9-glszm_GrayLevelVariance, 10-glszm_HighGrayLevelZoneEmphasis, 11-glszm_LargeAreaHighGrayLevelEmphasis, 12-glszm_LowGrayLevelZoneEmphasis, 13-gldm_DependenceNonUniformity, 14-gldm_GrayLevelNonUniformity, 15-gldm_LargeDependenceHighGrayLevelEmphasis, 16-gldm_SmallDependenceLowGrayLevelEmphasis, 17-ngtdm_Coarseness, 18-ngtdm_Contrast |
| Wavelet_HHL (11) | 1-firstorder_TotalEnergy, 2-glcm_Correlation, 3-glcm_Imc1, 4-glrlm_GrayLevelNonUniformity, 5-glrlm_RunLengthNonUniformity, 6-glszm_LargeAreaEmphasis, 7-glszm_LargeAreaHighGrayLevelEmphasis, 8-glszm_ZoneVariance, 9-gldm_DependenceNonUniformity, 10-gldm_GrayLevelNonUniformity, 11-ngtdm_Coarseness |
| Wavelet_HLH (28) | 1-firstorder_Energy, 2-firstorder_Kurtosis, 3-firstorder_Minimum, 4-firstorder_TotalEnergy, 5-glcm_Autocorrelation, 6-glcm_Idm, 7-glcm_Idn, 8-glcm_InverseVariance, 9-glcm_JointAverage, 10-glcm_SumAverage, 11-glrlm_HighGrayLevelRunEmphasis, 12-glrlm_LongRunHighGrayLevelEmphasis, 13-1glrlm_LowGrayLevelRunEmphasis, 14-glrlm_RunLengthNonUniformity, 15-glrlm_ShortRunLowGrayLevelEmphasis, 16-glszm_GrayLevelNonUniformity, 17-glszm_HighGrayLevelZoneEmphasis, 18-glszm_LowGrayLevelZoneEmphasis, 19-glszm_SizeZoneNonUniformity, 20-glszm_SmallAreaHighGrayLevelEmphasis, 21-glszm_SmallAreaLowGrayLevelEmphasis, 22-glszm_ZoneEntropy, 23-gldm_DependenceNonUniformity, 24-gldm_HighGrayLevelEmphasis, 25-gldm_LargeDependenceHighGrayLevelEmphasis, 26-gldm_LowGrayLevelEmphasis, 27-gldm_SmallDependenceLowGrayLevelEmphasis, 28-ngtdm_Coarseness |
| Wavelet_HLL (13) | 1-firstorder_TotalEnergy, 2-glcm_Imc1, 3-glcm_MCC, 4-glrlm_GrayLevelNonUniformity, 5-glrlm_RunLengthNonUniformity, 6-glszm_GrayLevelNonUniformity, 7-glszm_LargeAreaHighGrayLevelEmphasis, 8-glszm_ZoneEntropy, 9-gldm_DependenceNonUniformity, 10-gldm_GrayLevelNonUniformity, 11-ngtdm_Busyness, 12-ngtdm_Coarseness, 13-ngtdm_Strength |
| Wavelet_LHH (16 Features) | 1-firstorder_Kurtosis, 2-glcm_Idmn, 3-glcm_Idn, 4-glcm_InverseVariance, 5-glrlm_GrayLevelNonUniformity, 6-glrlm_RunLengthNonUniformity, 7-glszm_GrayLevelNonUniformity, 8-glszm_LargeAreaHighGrayLevelEmphasis, 9-gldm_DependenceNonUniformity, 10-gldm_DependenceNonUniformityNormalized, 11-gldm_DependenceVariance, 12-gldm_GrayLevelNonUniformity, 13-gldm_LargeDependenceHighGrayLevelEmphasis, 14-gldm_SmallDependenceLowGrayLevelEmphasis, 15-ngtdm_Coarseness, 16-ngtdm_Contrast |
| Wavelet_LHL (13 Features) | 1-firstorder_Kurtosis, 2-firstorder_Skewness, 3-firstorder_TotalEnergy, 4-glcm_Idn, 5-glrlm_GrayLevelNonUniformity, 6-glrlm_RunLengthNonUniformity, 7-glszm_GrayLevelNonUniformity, 8-glszm_LargeAreaEmphasis, 9-glszm_LargeAreaHighGrayLevelEmphasis, 10-glszm_ZoneVariance, 11-gldm_GrayLevelNonUniformity, 12-ngtdm_Coarseness, 13-ngtdm_Contrast |
| Wavelet_LLH (11) | 1-firstorder_Kurtosis, 2-firstorder_TotalEnergy, 3-glcm_Idmn, 4-glcm_Idn, 5-glrlm_GrayLevelNonUniformity, 6-glrlm_RunLengthNonUniformity, 7-glszm_GrayLevelNonUniformity, 8-gldm_DependenceNonUniformity, 9-gldm_GrayLevelNonUniformity, 10-gldm_SmallDependenceLowGrayLevelEmphasis, 11-ngtdm_Coarseness |
| Wavelet_LLL (7) | 1-glcm_Imc1, 2-glcm_MCC, 3-glrlm_GrayLevelNonUniformity, 4-glrlm_RunLengthNonUniformity, 5-glszm_GrayLevelNonUniformity, 6-gldm_GrayLevelNonUniformity, 7-ngtdm_Coarseness |
| LoG sigma-1-0-mm (11) | 1-firstorder_Kurtosis, 2-glcm_Imc1, 3-glcm_Imc2, 4-glcm_MCC, 5-glrlm_GrayLevelNonUniformity, 6-glrlm_RunLengthNonUniformity, 7-glszm_GrayLevelNonUniformity, 8-glszm_LargeAreaHighGrayLevelEmphasis, 9-gldm_DependenceNonUniformity, 10-gldm_GrayLevelNonUniformity, 11-ngtdm_Coarseness |
| LoG sigma-3-0-mm (15) | 1-firstorder_Kurtosis, 2-firstorder_TotalEnergy, 3-glcm_Imc1, 4-glrlm_GrayLevelNonUniformity, 5-glrlm_LongRunHighGrayLevelEmphasis, 6-glrlm_RunLengthNonUniformity, 7-glrlm_ShortRunLowGrayLevelEmphasis, 8-glszm_GrayLevelNonUniformity, 9-glszm_LargeAreaHighGrayLevelEmphasis, 10-gldm_DependenceNonUniformity, 11-gldm_GrayLevelNonUniformity, 12-gldm_LargeDependenceHighGrayLevelEmphasis, 13-1gldm_LowGrayLevelEmphasis, 14-gldm_SmallDependenceLowGrayLevelEmphasis, 15-ngtdm_Coarseness |
| Square (9) | 1-firstorder_TotalEnergy, 2-glcm_Imc1, 3-glrlm_GrayLevelNonUniformity, 4-glrlm_RunLengthNonUniformity, 5-glszm_GrayLevelNonUniformity, 6-glszm_LargeAreaHighGrayLevelEmphasis, 7-gldm_DependenceNonUniformity, 8-gldm_GrayLevelNonUniformity, 9-ngtdm_Coarseness |

Originally Refers to Conditions Where no Filter was Applied.

### LoG Filter

The Laplacian method searches for zero crossings in the second derivative of the image to find edges, but it has a disadvantage over blurred and noisy images. Therefore, first gaussian kernel is applied to the image then Laplacian is utilized for edge detection. The Gaussian kernel is used to smooth the image using Equation 4:

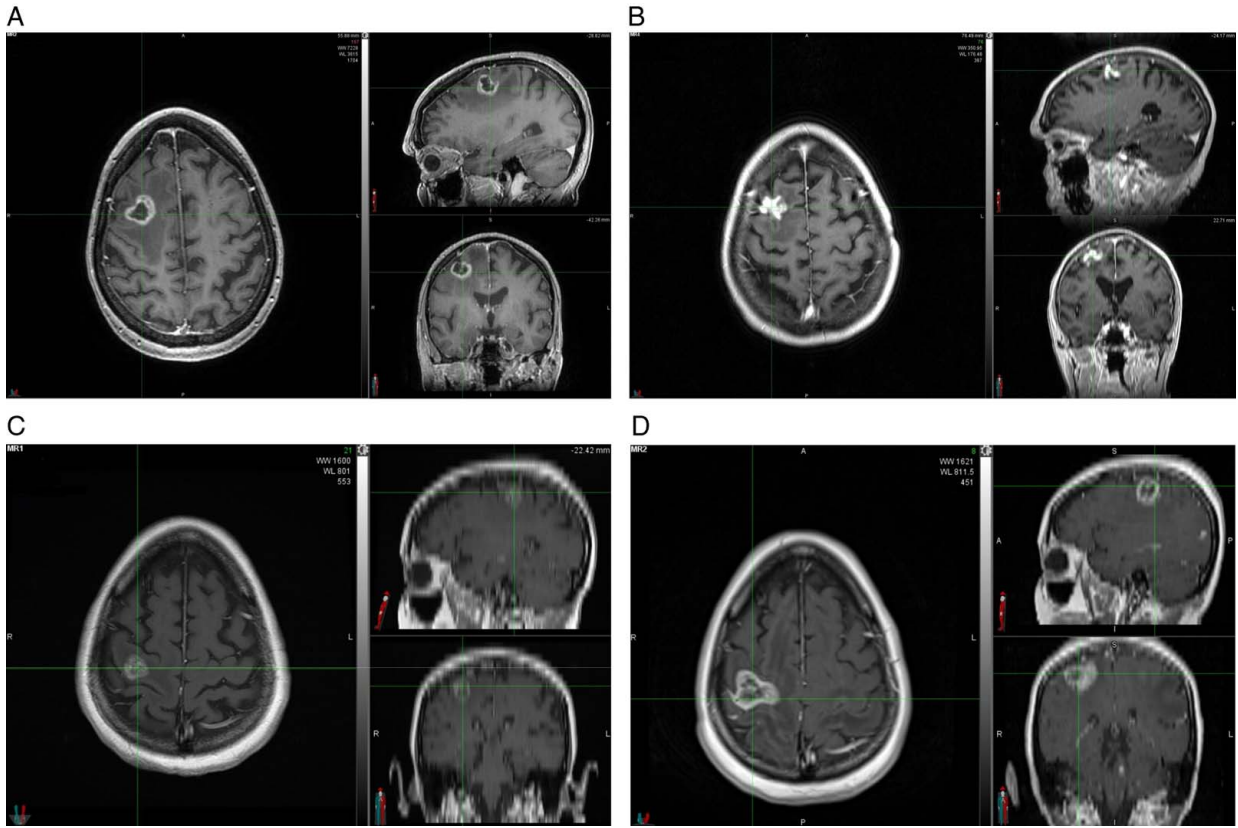

**FIGURE 1.** Progression and treatment effect. Radiation necrosis (A) 3 months, (B) 24 months. tumor recurrence, (C) 5 months, (D) 12 months after stereotactic radiosurgery. full color online

$$G(x, y, z, \sigma) = \frac{1}{(\sigma\sqrt{2\pi})^3} e^{-\frac{x^2+y^2+z^2}{2\sigma^2}} \quad \text{Eq. 4}$$

The width of the filter in the Gaussian kernel is defined by σ. Low sigma stress on fine structures (gray level changes over a short distance) and a high sigma emphasis on coarse structures (gray level change over a long distance). In the current study, the LoG filter was implemented using 2 kernel size values (σ = 1, and 3 mm) to include fine and coarse structures.

### Square Filter

The last filter implemented in the current study was Square which computes (Equation 5) the square of the image intensities. The resulting values are rescaled on the range of the initial original image.

$$F(x) = (cx)^2, \text{ where } c = \frac{\log(\max(x|)}{\max(x|)} \quad \text{Eq. 5}$$

Where x is the original intensity and F(x) is the filtered intensity.[18]

The square filter differs from other filters in that it is not an edge enhancement filter. Its purpose is to gauge the effectiveness of applying edge enhancement versus non-edge enhancement filters in radiomics analysis.

### Feature Extraction

Radiographic features from medical image data were extracted using PyRadiomics (Ver. 3.0.1) yielding 19 first-order statistics, 16 3D shape-based, 24 Gray Level Co-occurrence Matrix (GLCM), 16 Gray Level Run Length Matrix (GLRLM), 16 Gray Level Size Zone Matrix (GLSZM), 5 Neighboring Gray Tone Difference Matrix (NGTDM), and 14 Gray Level Dependence Matrix features (GLDM). As a result, 93 features included first, and second-order features which were computed and extracted per each filter. In total 1209 radiomics features were mined from the ROI of post-contrast T1-weighted for each patient in this study.

### Statistical Analysis and Feature Selection

The existence of too many features in the dataset not only extends the calculation time but also the absence of a semantic relationship between the features can reduce the classification accuracy. For this aim, the first normality test (Kolmogorov-Smirnov test) was performed using SPSS (Ver.27) for selecting parametric or non-parametric tests. Then, feature selection was applied by calculating the Mann-Whitney U Test for non-parametric data. For parametric data, an independent sample $t$ test was used between 2 groups of tumor recurrence and radiation necrosis.

### Classification

Machine-learning modeling was performed using the Python programming language version 3.7. First, data were shuffled and then randomly split into train/test sets (80%, 20%).

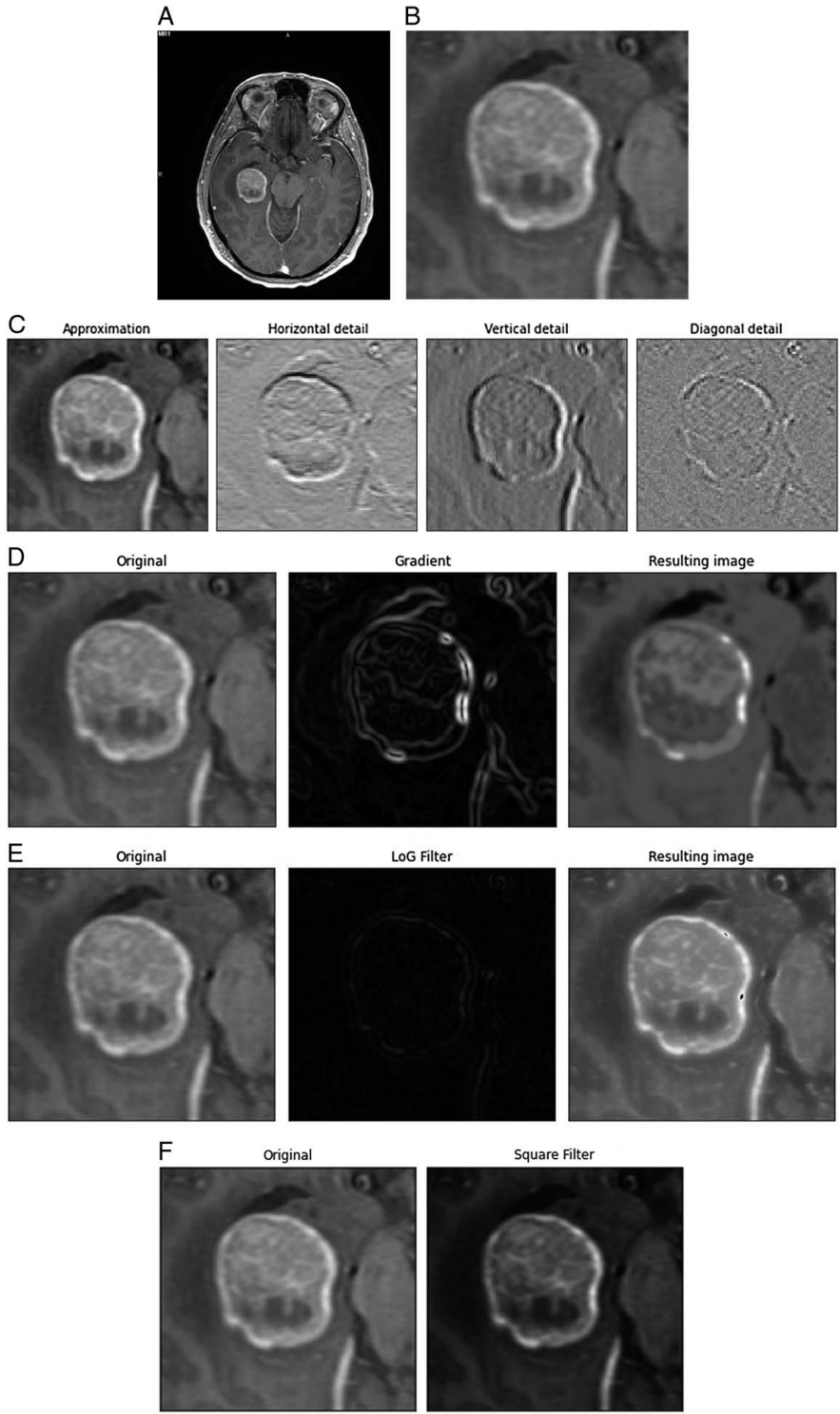

**FIGURE 2.** A, Original image of the right temporal tumor, (B) cropped image of right temporal tumor for better illustration. C, Two-dimensional discrete wavelet transform, (D) Gradient, (E) Laplacian-of-Gaussian, and (F) Square filter effect.

Because of the small dataset size, a 5-fold cross-validation was adopted and the classifiers were trained using the training fold only. Then standardization was implemented on feature values to rescale all data on a common scale (0,1). This is required due to some features falling between 0 and 1, whereas others fall in a very large range. Standardization gives equal weights/importance to each feature to eliminate bias due to individual features that have bigger values.

TABLE 2. (A) Average Classification Results After 5 Times Run for Gradient Filter, (B) LoG-Sigma1 and LoG-Sigma3, (C) Wavelet (HHH and& LLL), (D) Original and Square Filters After 5 Times Running

| Classifier | Filter | Accuracy | Specificity | Sensitivity | AUC |
|---|---|---|---|---|---|
| (A) | | | | | |
| RFC | Gradient | 0.80 ± 0.057 | 0.922 ± 0.059 | 0.696 ± 0.055 | 0.910 ± 0.047 |
| Logistic regression | Gradient | 0.718 ± 0.053 | 0.726 ± 0.087 | 0.812 ± 0.176 | 0.843 ± 0.082 |
| SVC | Gradient | 0.682 ± 0.088 | 0.665 ± 0.146 | 0.800 ± 0.274 | 0.78 ± 0.107 |
| (B) | | | | | |
| RFC | LoG sigma-1-0-mm | 0.755 ± 0.058 | 0.9206 ± 0.138 | 0.562 ± 0.162 | 0.906 ± 0.051 |
| RFC | LoG sigma-3-0-mm | 0.765 ± 0.091 | 0.919 ± 0.106 | 0.556 ± 0.115 | 0.869 ± 0.058 |
| Logistic regression | LoG sigma-1-0-mm | 0.729 ± 0.098 | 0.898 ± 0.057 | 0.476 ± 0.178 | 0.839 ± 0.050 |
| Logistic regression | LoG sigma-3-0-mm | 0.553 ± 0.095 | 0.481 ± 0.219 | 0.760 ± 0.208 | 0.739 ± 0.057 |
| SVC | LoG sigma-1-0-mm | 0.700 ± 0.059 | 0.903 ± 0.106 | 0.486 ± 0.228 | 0.855 ± 0.071 |
| SVC | LoG sigma-3-0-mm | 0.647 ± 0.136 | 0.669 ± 0.247 | 0.0696 ± 0.264 | 0.820 ± 0.026 |
| (C) | | | | | |
| RFC | Wavelet_HHH | 0.775 ± 0.038 | 0.957 ± 0.041 | 0.445 ± 0.136 | 0.887 ± 0.058 |
| RFC | Wavelet_LLL | 0.784 ± 0.071 | 0.922 ± 0.091 | 0.632 ± 0.112 | 0.859 ± 0.043 |
| Logistic regression | Wavelet_HHH | 0.753 ± 0.080 | 0.816 ± 0.123 | 0.816 ± 0.208 | 0.882 ± 0.051 |
| Logistic regression | Wavelet_LLL | 0.730 ± 0.084 | 0.921 ± 0.063 | 0.476 ± 0.160 | 0.823 ± 0.078 |
| SVC | Wavelet_HHH | 0.777 ± 0.062 | 0.850 ± 0.112 | 0.701 ± 0.084 | 0.890 ± 0.089 |
| SVC | Wavelet_LLL | 0.718 ± 0.080 | 0.856 ± 0.130 | 0.526 ± 0.147 | 0.767 ± 0.083 |
| (D) | | | | | |
| RFC | Original | 0.745 ± 0.060 | 0.876 ± 0.072 | 0.569 ± 0.207 | 0.827 ± 0.018 |
| RFC | Square | 0.686 ± 0.063 | 0.982 ± 0.079 | 0.294 ± 0.128 | 0.814 ± 0.032 |
| Logistic regression | Original | 0.718 ± 0.103 | 0.789 ± 0.143 | 0.654 ± 0.155 | 0.846 ± 0.080 |
| Logistic regression | Square | 0.730 ± 0.081 | 0.744 ± 0.147 | 0.812 ± 0.116 | 0.820 ± 0.076 |
| SVC | Original | 0.736 ± 0.058 | 0.833 ± 0.116 | 0.626 ± 0.180 | 0.828 ± 0.068 |
| SVC | Square | 0.729 ± 0.082 | 0.916 ± 0.085 | 0.447 ± 0.143 | 0.755 ± 0.078 |

AUC indicates area under the receiver operating characteristic curve; RFC, random forest classification; SVC, support vector classification.

Three different classifiers were adopted to demonstrate the chosen features' effectiveness. Logistic regression,[19] SVC,[20] and RFC,[21] classifiers were used to differentiate between RN and recurrence tumors. All 3 classifiers are discriminative models which separate classes and make predictions on the unseen data based on conditional probability. They fit the distribution of data to the binomial distribution and provide a category-related output with values between 1 and 2 (1 = tumor recurrence, and 2 = RN). GridSearchCV was implemented to tune the hyperparameters of each classification for each filter separately. To train logistic regression, the threshold was defined based on the difference between the true-positive rate (TPR) and the false-positive rate (FPR). The point at which the difference (TPR-FPR) is at its maximum value is the optimum value. The threshold was defined for each filter.

Each model was run 5 times on different train/test sets. The performance of each model was measured using the average of the following parameters: accuracy, the area under the curve (AUC), the value of the receiver operating characteristic curve, specificity, and sensitivity. The positive case for the confusion matrix was set to RN.

## RESULTS

According to the normalization test (Kolmogorov-Smirnov test) for all filters, Firstorder_Skewness, glcm_Correlation, glcm_Idn, glcm_Imc2, glszm_SmallAreaEmphasis, glszm_ZoneEntropy features were normally distributed ($P > 0.05$), whereas other features did not exhibit normal distributions ($P < 0.05$). The selected features resulted from the Mann-Whitney test for the nonparametric test and the independent sample $t$ test for the parametric test included first-order and second-order features and are tabulated in Table 1.

Figure 1 shows the progression and treatment effect of brain metastases on MRI. Both patients underwent the SRS technique (20 Gy in 1 fraction). Figure 1A and B demonstrate RN progression after 3 and 12 months of treatment. Figures 1C and D are images of a patient with true brain metastases progression after 5 and 12 months of radiation therapy.

Figure 2A shows the original image of the right temporal tumor before applying any filter and Figure 2B is the cropped section of the tumor for better visibility. Figures (2C-F) illustrate the effect of applying a different filter on the MR image.

Table 2 shows the results of all classifiers applying different filters. As we can see, the best diagnostic performance was obtained using RFC with a gradient filter.

The results of the confusion matrix for 5-fold cross-validation using Gradient filter for random forest classification, logistic regression, and SVC are shown in Figs. 3A, B, and C, respectively.

In Fig. 4. The receiver operating characteristic curve results for 5-fold cross-validation of 3 classifications used in this study (RFC logistic regression, and SVC) are displayed, respectively.

## DISCUSSION

Unfortunately, both brain metastasis and RN may have similar necrotic centers, irregular enhancing margins, and surrounding edema (Fig. 1). Currently, the gold standard for differentiating brain metastasis progression and RN is based on histopathological examination which is associated with procedure-related complications in 6% of cases.[22] Therefore, we explore a new method to differentiate between brain metastasis and RN based on comprehensive literature research on machine learning and radiomics analyses in neuroimaging for patients with brain metastases. Here we identified the optimal filter and machine learning (ML) model to classify brain metastasis progression and RN. The classification was performed based on a single conventional MRI post-contrast T1W image. Several studies were conducted based on multiparametric MRI data, including advanced imaging methods such as diffusion, perfusion, and MR spectroscopy to classify brain tumors.[23–26] However, advanced imaging is not available in all MR protocols and all sites; therefore, the ability to classify brain metastasis progression and RN was based on a single, commonly used sequence T1-weighted post-contrast images.

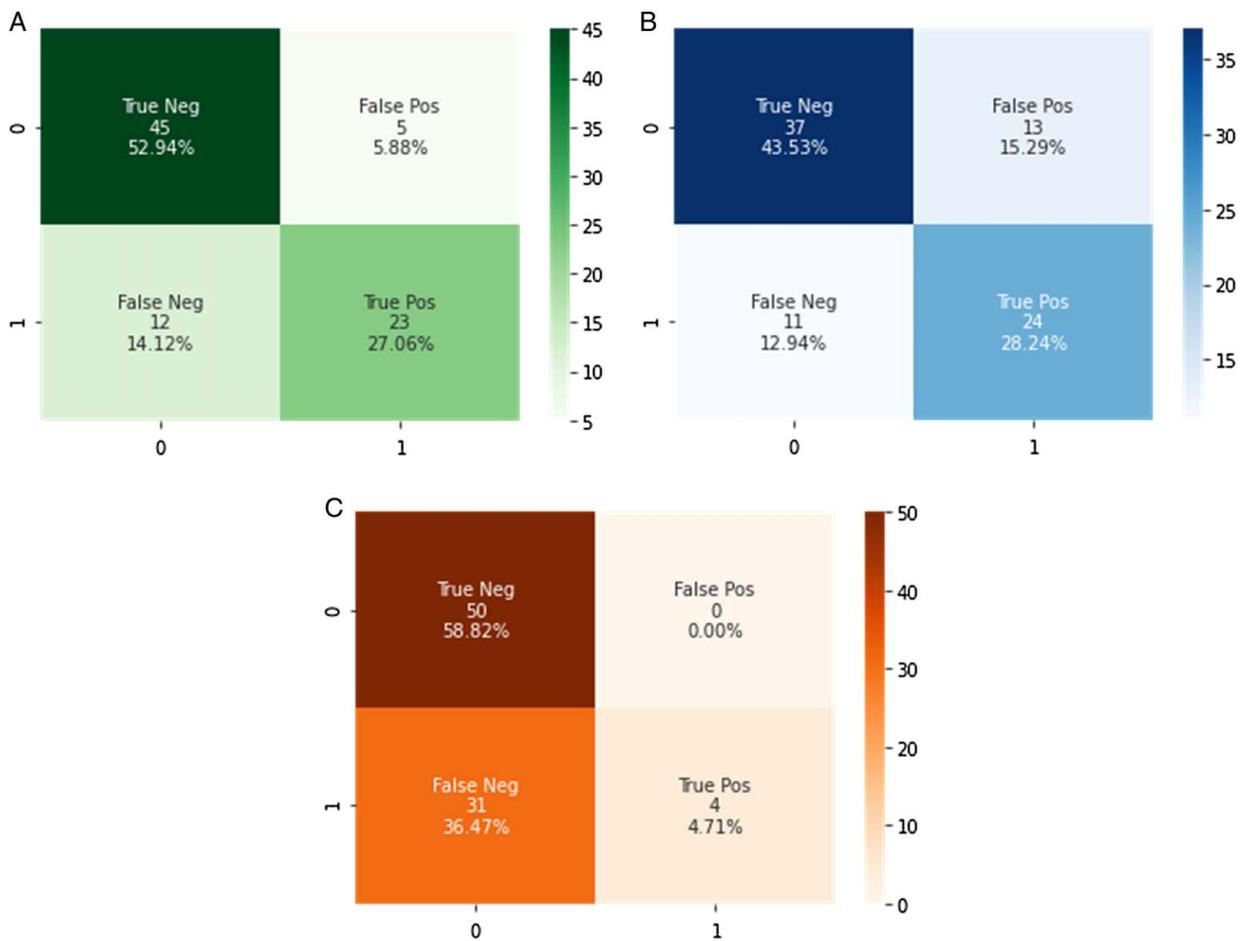

FIGURE 3. Confusion matrix result of 5-fold cross-validation using Gradient filter. A, random forest classification, (B) logistic regression, and (C) support vector classification. Neg indicates negative; Pos, positive.

Radiomics is an emerging field of medical image analysis that refers to the extraction of mineable data from medical imaging. It has been shown that applying radiomics in oncology improves diagnosis, prognostication, and clinical decision support, to deliver precision medicine.[12,27] The features which contributed to the classification between the 2 groups were clinical information, first-order and second-order radiomics features. In this way, we conducted a comprehensive characteristic of the tumor area including both micro and macro-level characteristics of the target. Furthermore, the majority of these features such as second-order, and some of the first-order features cannot be identified and quantified by the naked eye. In the current study, 1209 features per patient were extracted initially, followed by feature selection to identify a subset of features that are most likely to differentiate between the 2 groups and to reduce overfitting.[28] This process is important because it is difficult to classify RN and brain metastasis progression with high accuracy since their radiomic features are similar. Furthermore, success in radiomics depends on the type of data used. Consequently, feature selection was performed using the Mann-Whitney test and independent sample $t$ test for non-parametric and parametric data with SPSS (Ver.27), respectively.

Previous researchers have presented different automated approaches for the classification of brain tumors (primary and secondary). In 2017, Mohsen et al[29] used a deep neural network (DNN) to classify a dataset of 66 real brain MRIs (22 normal and 44 abnormal images) into 4 classes of normal, glioblastoma (GBM), sarcoma, and metastatic bronchogenic carcinoma tumors. The highest accuracy in terms of AUC was obtained using DNN. They concluded that using DTW used with DNN can provide good results. Artzi et al[12] researched data from 439 patients including 212 GBM and 227 brain metastases originating from breast, lung, and other cancers. They classified data using various machine-learning algorithms including support vector machine (SVM), k-nearest neighbor, decision trees, and ensemble classifiers. The SVM algorithm resulted in the best classification between all groups. They concluded that other classifications may perform better on other problems since there is no universally accepted optimal learning algorithm. Çinarer et al[10] used radiomics with DTW for classifying GBM grades (Grade II-III). For this aim, they evaluated data from 121 patients with different grades of GBM (Grade II, n = 77; Grade III, n = 44). For each patient, 744 radiomics features were extracted by applying low sub-band and high sub-band 3D DTW. Their study demonstrated that the right combination of radiomic analysis and deep learning methods can result in accurate prediction. In 2021, Park et al[30] established a radiomics strategy with machine learning from conventional and diffusion MRI to differentiate recurrent GBM from RN after concurrent chemoradiotherapy or radiotherapy. The study was conducted on 127 patients (86 recurrent GBM and 41 RN), and 263 radiomics features from conventional MRI sequences (T2-weighted and post-contrast T1-weighted images) and Apparent Diffusion Coefficient were extracted for analysis. They concluded using radiomics features extracted from Apparent Diffusion Coefficient may help differentiate recurrent GBM from RN. Peng et al[31] performed a study on 82 brain lesions (52 recurrent and 30 RN) using IsoSVM and logistic regression models. The highest AUC 0.81 and 0.7 was achieved for IsoSVm and logistic regression, respectively. Another study was conducted by Zhang et al[32] in 2018 to distinguish tumor progression from RN based on radiomic features from MRI. They retrospectively identified 87 patients with pathologically confirmed necrosis (24 lesions) or progression (73 lesions) and calculated 285 radiomic features from 4 MR sequences (T1, T1 post-contrast, T2, and fluid-attenuated inversion recovery) obtained at 2 follow-up time points per lesion per patient. The highest AUC of 0.73 for the RUSBoost classifier was obtained. Tiwari et al[33] used 58 patients to determine the feasibility of radiomic features in differentiating radiation necrosis from recurrent brain tumors on routine MR imaging (gadolinium T1WI, T2WI, FLAIR). The best-performing feature sets in distinguishing RN and tumor recurrence was obtained for FLAIR, with reported AUC 0f $0.79 \pm 0.05$. All results from previous studies suggest that radiomic analysis on routinely acquired MR imaging might help discriminate RN and tumor recurrence in brain cancer.

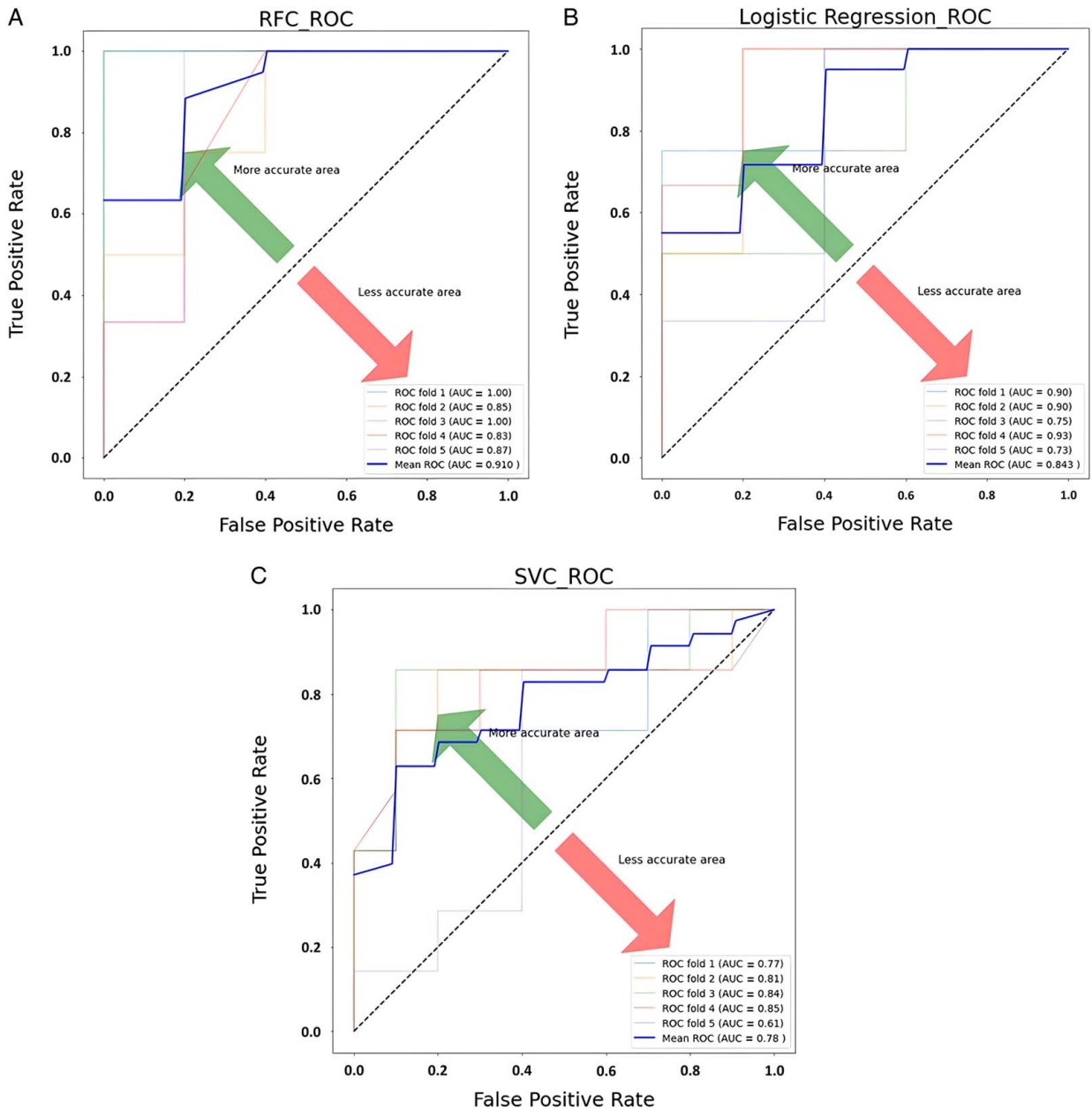

**FIGURE 4.** Receiver operating characteristic curve result of 5-fold cross-validation for 3 classifications (A) random forest classification, (B) logistic regression, and (C) support vector classification. AUC indicates area under the curve; RFC, random forest classification; ROC, receiver operating characteristic; SVC, support vector classification.

According to our literature review, there is no study to evaluate the effect of different filters such as DTW, gradient, Log sigma, and square on differentiating RN and brain metastasis progression using radiomics. Therefore, no data are available for comparing our results against existing published literature. Furthermore, RN is the late delayed radiation damage that may occur 6 months to years after radiation treatment.[26] Hence, we used at least 4 follow-up images with 3-month time intervals with additional clinical data such as PET scan, MRS, or pathology lab report to confirm necrosis and progression.

Our radiomics approach was tested with 3 classifications, and the highest average AUC value of 0.91 was achieved using a gradient filter with RF classification (Fig. 4A). For SVC the best results were obtained by using wavelet_HHH with a high average AUC of 0.890 (Table 2C). Logistic regression achieved the lowest average AUC compared with other classifiers with AUC = 0.882 using wavelet_HHH (Table 2C). These results indicate that the classifiers were successful at classifying the test cohort and the RFC was significantly better than other algorithms.

Our result showed different combinations of filters and ML models result in varying levels of agreement for the same criteria (Tables 2A-D). Applying square filter results in poor outcomes compared with the original data. Furthermore, the brain has fine tissue; therefore, wavelet_HHH and LoG-sigma1 provide better outcomes compared with wavelet_LLL and Log-sigma3 which are usually applied for coarse structures (Table 2). In particular, the RFC showed the best performance by achieving the highest AUC among other models. Both Logistic regression and SVC have a serious drawback, which is sensitive to outliers in training samples, and this reduces their generalization ability. Generally, Logistic regression and SVC are useful in differentiating groups into 2 different classes, but both perform well when the training data is less, and there are a large number of features.

These findings emphasize the importance of understanding the response and limitations of each filter along with the ML model. Moreover, our data demonstrated using the right radiomics features coupled with advanced machine-learning algorithms can improve differentiation between brain metastasis progression and RN even without applying filters (Table 2).

This retrospective study highlights the advantage of using automatic methods and extraction of high-order statistical features to assist in radiological assessment and clinical decision-making. However, we acknowledge this study has limitations such as using a small sample of data due to RN being the late effect of radiotherapy and unfortunately few patients survived to develop RN. Also, using only MR post-contrast T1W images and including different acquisition parameters may affect the outcome. This can be resolved by having a large and mixed dataset. Moreover, splitting this group into subgroups according to the main tumor origin, and substantially increasing the number of patients in each category, may significantly improve classification outputs. Moreover, the research concentrated solely on distinguishing between RN and genuine progression, without taking into account other clinical aspects like patient characteristics or treatment history that could impact the models' accuracy. Subsequent studies could delve into these factors and their influence on radiomics-based predictive models.

## CONCLUSION

The outcome of this study indicates using radiomics with machine-learning models holds great promise for discriminating between RN and true progression in brain metastases. In a comparison of different filters, it is shown that edge enhancement filters can improve prediction performance. Moreover, among all edge detection filters LoG sigma 1 and wavelet_HHH provide better outcomes. This can be attributed to the fine structure of the brain. RFC with gradient filter provided the best prediction performance. Logistic regression and SVC showed the best diagnostic performance using DWT with a high-pass filter. In conclusion, different filter and ML model combinations showed AUCs ranging from 0.739 to 0.918. These findings emphasize the importance of understanding the response and limitations of each filter along with the ML model. We believe this approach maintains its effectiveness in the clinical approach, as it is a noninvasive technique.

## ACKNOWLEDGMENTS

*The authors express their sincere thanks for the useful discussions with Kevin Shuai Xu, PhD (Case Western Reserve University) and Kerry Krugh, PhD (ProMedica Health System).*